\documentclass[a4paper, conference]{IEEEtran}
\usepackage{cite}
\usepackage{amsmath,amssymb,amsfonts}
\usepackage{algorithmic}
\usepackage{graphicx}
\usepackage{textcomp}
\usepackage{xcolor}
\usepackage{tabularray}
\usepackage{booktabs} 

\usepackage{pifont} 
\newcommand{\cmark}{\ding{51}}
\newcommand{\xmark}{\ding{55}}
\usepackage{soul}   
\usepackage{etoolbox}
\newcommand{\revref}[1]{\colorbox{yellow}{\cite{#1}}}
\robustify{\revref}
\usepackage{threeparttable} 

\usepackage{background}
\usepackage{xcolor}
\usepackage{hyperref}

\backgroundsetup{
  scale=1,
  color=black,
  opacity=1,
  angle=0,
  position=current page.south,
  vshift=10pt,
  contents={\textcolor{red}{© 2025. For personal use only. Published version under doi:  \href{https://doi.org/10.23919/SoftCOM66362.2025.11197453}{10.23919/SoftCOM66362.2025.11197453}. }}
}

\begin{document}
\title{Toward Secure Content-Centric Approaches for 5G-Based IoT: Advances and Emerging Trends}

\author{
  \IEEEauthorblockN{Ghada Jaber, Mohamed Ali Zormati, Walid Cavelius, Louka Chapiro, Mohamed El Ahmadi}
  \IEEEauthorblockA{
    Université de Technologie de Compiègne, Sorbonne Universités, CNRS,\\
    Heudiasyc UMR 7253, CS 60 319 - 60 203 Compiègne Cedex, France\\
    Email: ghada.jaber@hds.utc.fr, mohamed-ali.zormati@hds.utc.fr,\\
    walid.cavelius@etu.utc.fr, louka.chapiro@etu.utc.fr, mohamed.el-ahmadi@etu.utc.fr
  }
}

\maketitle

\begin{abstract}
The convergence of the Internet of Things (IoT) and 5G technologies is transforming modern communication systems by enabling massive connectivity, low latency, and high-speed data transmission. In this evolving landscape, Content-Centric Networking (CCN) is emerging as a promising alternative to traditional Internet Protocol (IP)-based architectures. CCN offers advantages such as in-network caching, scalability, and efficient content dissemination, all of which are particularly well-suited to the constraints of the IoT. However, deploying content-centric approaches in 5G-based IoT environments introduces significant security challenges. Key concerns include content authentication, data integrity, privacy protection, and resilience against attacks such as spoofing and cache poisoning. Such issues are exacerbated by the distributed, mobile, and heterogeneous nature of IoT and 5G systems. In this survey, we review and classify existing security solutions for content-centric architectures in IoT-5G scenarios. We highlight current trends, identify limitations in existing approaches, and outline future research directions with a focus on lightweight and adaptive security mechanisms.

\end{abstract}


\section{Introduction}
The rapid growth of the Internet of Things (IoT) and the widespread deployment of 5G networks are profoundly transforming communication paradigms. IoT applications, ranging from smart cities and autonomous vehicles to industrial automation and healthcare systems, demand not only ubiquitous connectivity but also efficient, secure, and scalable data management \cite{rahmaniInternetThingsApplications2022}. At the same time, 5G technologies bring new capabilities — such as ultra-reliable low-latency communication (URLLC), enhanced mobile broadband (eMBB), and massive machine-type communication (mMTC) — that are essential for meeting the diverse requirements of next-generation IoT systems \cite{wijethilakaSurveyNetworkSlicing2021}.

Given the evolving demands of IoT applications and the advanced capabilities introduced by 5G networks, Content-Centric Networking (CCN) has emerged as a compelling alternative to the traditional host-centric Internet Protocol (IP) communication model \cite{c.n.SystematicSurveyContent2023}. Rather than focusing on the location of data, CCN centers communication on the content itself. This paradigm shift enables key features such as in-network caching, native multicast, and content-based security, making it particularly well-suited to the dynamic, resource-constrained, and mobility-intensive conditions that characterize IoT-5G environments \cite{chenContentcentricFrameworkInternet2022}.

However, despite its conceptual advantages, integrating content-centric architectures into 5G-based IoT systems introduces several security challenges \cite{nourInformationCentricNetworkingWireless2021}. Traditional security models are often location-dependent or centralized, which makes them poorly suited to the decentralized nature of CCN. Furthermore, IoT devices typically have limited processing power, memory, and energy, which complicates the deployment of conventional security mechanisms. Threats such as content poisoning, interest flooding, spoofing, and unauthorized data access are more critical in scenarios with high mobility, heterogeneous devices, and intermittent connections, which are common in 5G-enabled IoT \cite{shurSEDIMENTIoTdevicecentricMethodology2022}.

These challenges raise a critical research question: How can security be effectively and efficiently enforced in content-centric architectures deployed within 5G-enabled IoT, given the constraints and vulnerabilities inherent to both technologies? To address this question, this survey provides an overview of existing approaches and techniques aimed at securing CCN in IoT-5G applications. It categorizes the state-of-the-art solutions, highlights their strengths and limitations, and identifies open issues that remain unresolved. Special attention is given to emerging trends such as lightweight cryptographic mechanisms, trust management models, and the use of machine learning for adaptive threat detection \cite{anjumNamedDataNetworking2024}.

There has been a significant research effort exploring the individual and combined roles of the IoT, 5G, and CCN, particularly with consideration of their security aspects. However, to the best of our knowledge, no existing work has focused on the security of content-centric 5G-based IoT. Thus, this work aims to address this gap and serve as a starting point for researchers and practitioners designing secure content-centric solutions for next-generation IoT-5G systems.

The remainder of the paper is organized as follows. Section II introduces the background of IoT, 5G, and CCN technologies. Section III highlights the key security requirements of CCN for 5G-based IoT. Section IV provides an overview of recent advances in security solutions for content-centric 5G-based IoT. Emerging trends and the most prominent future directions are discussed in Section V. Finally, the review is concluded in Section VI.

\section{Background}
In this section, we present the main concepts of IoT, 5G networks, and CCN, highlighting their core capabilities and how their integration opens up new opportunities.

\subsection{Internet of Things (IoT)}
The Internet of Things (IoT) refers to a distributed network that connects billions of objects (i.e., things) to each other and to the Internet. This connectivity enables a wide range of services across various domains \cite{zormatiOverviewMachineLearningEnabled2023}. Notable examples include smart city deployments, connected healthcare systems, and smart agriculture solutions.

The complexity of the IoT primarily stems from the tremendous number of heterogeneous devices connected to the Internet and communicating together. These devices often operate in constrained environments where computational and energy efficiency are critical \cite{maoEnergyEfficientIndustrialInternet2021}. Such limitations pose significant security challenges because conventional security mechanisms tend to be resource-intensive. Nevertheless, real-world IoT implementations must ensure essential security properties such as integrity, authentication, and confidentiality \cite{rachitSecurityTrendsInternet2021}.

To effectively support its diverse applications, the IoT relies on several networking technologies, most notably the fifth generation of telecommunications networks, known as 5G. 5G is widely recognized as a key enabler of large-scale IoT deployment \cite{wijethilakaSurveyNetworkSlicing2021}.

\subsection{Fifth-Generation (5G) Networks}
The fifth-generation (5G) mobile communication system, defined by the 3rd Generation Partnership Project (3GPP), represents a major advancement in mobile telecommunications. It introduces significant improvements in data rates, latency, reliability, and support for massive device connectivity \cite{khanhWirelessCommunicationTechnologies2022}. The 5G requirements and vision for applications and services are captured in three primary usage scenarios: eMBB, URLLC, and mMTC. The relevance of specific key capabilities depends on the use case, although all key capabilities are significant in most cases. All of these scenarios are relevant to IoT use cases and are defined as follows \cite{erunkulu5GMobileCommunication2021}.

\begin{itemize}
    \item \textbf{Enhanced Mobile Broadband (eMBB):} Targets high data throughput and extended coverage for bandwidth-intensive applications. It enables IoT use cases such as augmented and virtual reality, remote diagnostics, etc.
    \item \textbf{Ultra-Reliable Low Latency Communications (URLLC):} Focuses on delivering ultra-low latency (below 1 ms) and extremely high reliability (99.999\%) for mission-critical services. This is essential for time-sensitive IoT applications like autonomous driving, industrial automation, and remote surgery.
    \item \textbf{Massive Machine-Type Communications (mMTC):} Supports the simultaneous connection of a vast number of low-power devices (up to one million per square kilometer) making it ideal for large-scale IoT deployments such as smart cities, precision agriculture, and industrial monitoring.
\end{itemize}

By addressing the diverse communication requirements, 5G serves as a foundational enabler for the large-scale deployment of IoT systems. However, despite its advances in connectivity and performance, 5G has traditionally relied on a host-centric communication model, which limits performance in highly dynamic, data-intensive environments. Recent developments, particularly through the Open Radio Access Network (O-RAN) architecture, introduce disaggregated control via elements such as the RAN Intelligent Controller (RIC), enabling more flexible, intelligent, and service-oriented paradigms that move beyond strict host-centricity. To further address these challenges, content-oriented architectures have emerged as a promising alternative, offering greater efficiency, scalability, and flexibility for next-generation IoT-5G ecosystems \cite{zhangInNetworkCachingICNBased2023}.

\subsection{Content-Centric Networking (CCN)}
Content-Centric Networking (CCN) represents a paradigm shift from the traditional host-based IP model toward a content-oriented communication architecture. In CCN, data is requested and delivered based on content names rather than host addresses, allowing users to retrieve information regardless of its physical location \cite{jaberCollaborativeCachingStrategy2020a}. In CCN, communication occurs through the exchange of two packet types: Interest packets, which request content by name, and Data packets, which carry the requested content and its signature \cite{jacobson2009ccn}. 

This design enables several key features that are particularly relevant for dynamic, distributed, and resource-constrained IoT-5G systems \cite{zhangInNetworkCachingICNBased2023}:
\begin{itemize}
    \item \textbf{In-network caching:} Intermediate nodes can temporarily cache content, enabling future requests to be served locally. This reduces latency, minimizes bandwidth usage, and improves data availability. This is valuable for IoT devices that typically have limited connectivity.
    \item \textbf{Native multicast support:} Any node that holds the matching data can satisfy a single interest (i.e., request) packet, enabling efficient one-to-many data distribution without redundant transmissions. This is particularly beneficial for IoT deployments (e.g., firmware updates).
    \item \textbf{Decoupling of content from location:} Data retrieval is based on content names rather than network addresses. This approach supports mobility and dynamic topologies. This flexibility is essential for IoT-5G systems, as devices often change locations or network conditions.
\end{itemize}

CCN provides valuable security features, one of which is content-based security \cite{jaberAdaptiveDutycycleMechanism2020}. In this design, security is embedded directly into the data, typically via cryptographic signatures. This ensures the integrity and authenticity of the data, regardless of the source or delivery path. 

Despite its advantages, integrating CCN into IoT-5G environments presents security challenges \cite{nourInformationCentricNetworkingWireless2021}. CCN lacks native mechanisms for confidentiality, access control, and privacy. Cached data in the network is difficult to restrict to authorized users, and the open, distributed nature of in-network caching exposes the system to threats such as unauthorized access, interest flooding, and content poisoning.

These issues are amplified by the limited resources of IoT devices and the high degree of mobility and heterogeneity of 5G networks. Therefore, designing a secure content-centric 5G-based IoT requires addressing a set of security requirements with novel, lightweight, context-aware mechanisms that go beyond conventional approaches.

\section{Security Requirements in Content-Centric 5G-Based IoT}
The combination of CCN, 5G networks, and the IoT creates a powerful yet challenging environment. Each paradigm offers its own advantages and presents different challenges. Ensuring security in this multifaceted environment requires addressing the vulnerabilities inherent to each technology. In the following, we introduce the key security requirements for content-centric 5G-based IoT.

\begin{itemize}
    \item \textbf{Resource constraints of IoT devices:} IoT devices typically have severe limitations in terms of processing power, energy, and memory. Traditional cryptographic solutions, such as Advanced Encryption Standard (AES) or Rivest-Shamir-Adleman (RSA), are too computationally expensive \cite{ranaLightweightCryptographyIoT2022}. Therefore, lightweight cryptographic schemes are necessary.
    \item \textbf{Confidentiality and integrity in 5G:} 5G networks introduce security enhancements, such as the use of temporary identifiers and robust key distribution schemes, to preserve user confidentiality. However, vulnerabilities can still arise from unreliable hardware or misconfigured network slicing policies \cite{ahmedSecure5GEnabledInternet2024}. The large-scale wireless nature of 5G increases exposure to risks such as eavesdropping and signal jamming, and rogue base stations \cite{sullivan5GSecurityChallenges2021}, all of which must be considered when designing secure IoT systems.
    \item \textbf{Content integrity and authenticity in CCN:} In CCN, data is cached and retrieved based on content names. This model requires mechanisms that ensure content has not been altered during caching or transmission. Solutions include digital signatures, Hash-Based Message Authentication Codes (HMACs), and secure hash-based naming \cite{ullahLightweightIdentityBasedSignature2020}. However, due to limitations of IoT devices, these verification methods must be efficient.
    \item \textbf{Access control and trust management:} Since CCN decouples content from its source, access control must be enforced at the content level rather than through traditional, host-based mechanisms. This requires dynamic, context-aware access control schemes to manage permissions across distributed caches. Various trust management strategies have been proposed, including centralized trusted authorities, blockchain-based frameworks, zero-trust architectures, and named tokens \cite{liFutureIndustryInternet2024}.
    \item \textbf{Attack detection and resilience:} The convergence of the IoT, 5G, and CCN technologies has created a complex threat landscape that is vulnerable to Distributed Denial-of-Service (DDoS) attacks, content poisoning, spoofing, and unauthorized data access, to name a few. Edge-based anomaly detection mechanisms that leverage machine learning (ML) models have proven effective in combating these threats \cite{afaqMachineLearning5G2021}. Multi-access edge computing (MEC) further enhances security by enabling localized threat detection and mitigation, improving real-time responsiveness, and reducing centralized bottlenecks \cite{ranaweeraMECenabled5GUse2022}.
\end{itemize}

Securing content-centric, 5G-based IoT systems demands a cross-layered approach. Resource-constrained IoT devices require lightweight cryptography, while 5G capabilities can be used to isolate and delegate security tasks. CCN needs embedded, data-centric protections such as content-naming-based signatures and decentralized trust. Effective defense combines secure transmission, adaptive access control, distributed trust management, and edge-level threat detection to build scalable, resilient, and trustworthy systems at the intersection of IoT, 5G, and CCN.

\section{Review of Existing Security Solutions}
This section provides an overview of existing security solutions in the context of the integration of IoT, 5G, and CCN, and evaluates how these solutions address the key challenges outlined beforehand. The solutions are categorized into three main areas: cryptographic mechanisms, trust management models, and access control with privacy preservation. This structure reflects the multifaceted nature of securing content-centric 5G-based IoT systems.

\begin{table*}[ht]
\centering
\caption{Overview of Recent Security Solutions for Content-Centric 5G-Based IoT Systems}
\label{tab:overview_solutions}
\begin{tblr}{
  row{2} = {c},
  column{2} = {c},
  column{3} = {c},
  column{4} = {c},
  column{5} = {c},
  column{6} = {c},
  cell{1}{1} = {r=2}{},
  cell{1}{2} = {r=2}{},
  cell{1}{3} = {r=2}{},
  cell{1}{4} = {r=2}{},
  cell{1}{5} = {r=2}{},
  cell{1}{6} = {r=2}{},
  cell{1}{7} = {c=5}{c},
  cell{3}{7} = {c},
  cell{3}{8} = {c},
  cell{3}{9} = {c},
  cell{3}{10} = {c},
  cell{3}{11} = {c},
  cell{4}{7} = {c},
  cell{4}{8} = {c},
  cell{4}{9} = {c},
  cell{4}{10} = {c},
  cell{4}{11} = {c},
  cell{5}{7} = {c},
  cell{5}{8} = {c},
  cell{5}{9} = {c},
  cell{5}{10} = {c},
  cell{5}{11} = {c},
  cell{6}{7} = {c},
  cell{6}{8} = {c},
  cell{6}{9} = {c},
  cell{6}{10} = {c},
  cell{6}{11} = {c},
  cell{7}{7} = {c},
  cell{7}{8} = {c},
  cell{7}{9} = {c},
  cell{7}{10} = {c},
  cell{7}{11} = {c},
  cell{8}{7} = {c},
  cell{8}{8} = {c},
  cell{8}{9} = {c},
  cell{8}{10} = {c},
  cell{8}{11} = {c},
  cell{9}{7} = {c},
  cell{9}{8} = {c},
  cell{9}{9} = {c},
  cell{9}{10} = {c},
  cell{9}{11} = {c},
  cell{10}{7} = {c},
  cell{10}{8} = {c},
  cell{10}{9} = {c},
  cell{10}{10} = {c},
  cell{10}{11} = {c},
  hline{1,11} = {-}{0.08em},
  hline{2} = {7-11}{},
  hline{3-10} = {-}{},
}
\textbf{Reference }        & \textbf{Year } & \textbf{IoT }         & \textbf{5G }          & \textbf{CCN }         & \textbf{Lightweight } & \textbf{Security Requirements} &                          &                           &                       &                       \\
                           &                &                       &                       &                       &                       & \textbf{Integrity}             & \textbf{Confidentiality} & \textbf{Authentification} & \textbf{Control}      & \textbf{Trust}        \\
\cite{ullahLightweightIdentityBasedSignature2020} & 2020           & \cmark & \xmark & \cmark & \cmark & \cmark          & \xmark    & \cmark     & \xmark & \xmark \\
\cite{hussainNovelEfficientCertificateless2021} & 2021           & \cmark & \xmark & \cmark & \cmark & \cmark          & \xmark    & \cmark     & \xmark & \xmark \\
\cite{singhNDNContentPoisoning2021} & 2021           & \xmark & \xmark & \cmark & \cmark & \cmark          & \xmark    & \xmark     & \xmark & \cmark \\
\cite{nourInformationCentricNetworkingWireless2021} & 2022           & \xmark & \cmark & \cmark & \xmark & \xmark          & \xmark    & \cmark     & \xmark & \xmark \\
\cite{srikanthProxyBasedReEncryptionDesign2023} & 2023           & \cmark & \xmark & \cmark & \cmark & \xmark          & \cmark    & \xmark     & \cmark & \xmark \\
\cite{mecerhedEfficientDecentralizedFinegrained2024} & 2024           & \cmark & \xmark & \cmark & \cmark & \xmark          & \cmark    & \cmark     & \cmark & \xmark \\
\cite{wangAnonymousEfficientCertificateless2024} & 2024           & \cmark & \xmark & \cmark & \cmark & \cmark          & \xmark    & \cmark     & \xmark & \cmark \\
\cite{sahaECMHPECCBasedSecure2024} & 2024           & \cmark & \xmark & \cmark & \cmark & \cmark          & \cmark    & \xmark     & \xmark & \xmark \\

\end{tblr}
\begin{tablenotes}
\small
\item A (\cmark{}) denotes that the corresponding aspect is addressed in the work; a (\xmark{}) indicates it is not.
\end{tablenotes}
\end{table*}

\subsection{Cryptographic Mechanisms}
Cryptographic mechanisms form the foundation of all subsequent security services. These mechanisms enable data confidentiality through encryption, data authenticity and integrity through digital signatures, and secure key exchanges, which are essential for secure communication.

Asymmetric cryptography allows content producers to sign data for widespread verification of its authenticity and integrity. Examples of these algorithms include RSA and Elliptic Curve Cryptography (ECC), which are often used in identity-based schemes \cite{shaabanEfficientECCbasedAuthentication2023}. However, public key operations can be computationally intensive for resource-constrained IoT devices, as well as for network-wide verification. The use of Public Key Infrastructure (PKI) also introduces additional computational overhead.

Symmetric key cryptography is generally faster for encrypting data once a shared key is established, making it suitable for constrained devices if key distribution is managed efficiently. Schemes based on symmetric principles, such as those using hash functions and \textit{XOR} operations for token generation, are designed to be lightweight and efficient. Although robust algorithms like AES exist, their suitability for the most constrained devices is still a concern. Consequently, a hybrid approach using asymmetric methods for key establishment and symmetric methods for data protection is often practical. For example, the Elliptic Curve based Multicasting Handshake Protocol (ECMHP) \cite{sahaECMHPECCBasedSecure2024} proposes an ECC-based secure handshake mechanism for multicasting in CCN-IoT environments to efficiently secure group communications.

In CCN systems, security is bound to the data itself. Content-based signatures attach digital signatures to data packets to verify their authenticity and integrity network-wide. This is key to preventing attacks like content poisoning, in which malicious content is injected into the network. Producers sign the content, and these signatures travel with the data, enabling any node to verify them. The lightweight identity-based signature (IBS) scheme using HECC, as proposed in \cite{ullahLightweightIdentityBasedSignature2020}, is an example of a solution tailored for IoT-based CCN. It aims to reduce PKI overhead while ensuring content integrity and authenticity.

Additionally, the limited resources of IoT devices make lightweight cryptography essential. Traditional algorithms are often too demanding. There is a strong focus on developing cryptographic mechanisms, including ciphers, hash functions, and signature schemes, that provide adequate security while minimizing computational and communication costs.

\subsection{Trust Management Models}
Using cryptographic mechanisms naturally presents the challenge of managing trust within the network. Various approaches have been proposed, which we categorize into two main groups: those that rely on a centralized trust entity and those that implement decentralized trust architectures.

Cryptography, especially digital signatures, effectively guarantees content integrity and authenticity, thus protecting against attacks such as content poisoning. Centralized solutions, as in \cite{ullahLightweightIdentityBasedSignature2020} and \cite{hussainNovelEfficientCertificateless2021}, use trusted entities, such as a network manager or private key generator (PKG), to generate and distribute cryptographic keys. These approaches offer lightweight, identity-based signature schemes that minimize the computational burden of signature verification, providing a practical alternative to traditional PKI models.

An alternative strategy is to eliminate the need for a trusted authority by implementing a decentralized architecture. This can be achieved by setting up a reputation-based trust system. In a CCN network, any node can cache data and respond to an interest. Assigning a reputation level to nodes indicates the reliability of the content they relay.

Decentralized trust models aim to eliminate reliance on a central authority by leveraging reputation-based systems. In CCN environments, where any node may cache and serve data, assigning reputation levels to nodes reflects their reliability in content delivery. For example, the fuzzy reputation trust model proposed in \cite{singhNDNContentPoisoning2021} offloads signature verification to clients, which reduces the computational demands on intermediate nodes. This approach is well-suited for resource-constrained IoT devices. Additionally, blockchain technology has been explored to establish decentralized, certificateless trust frameworks that ensure content integrity and authenticity without the probabilistic uncertainties inherent in reputation systems, as demonstrated in \cite{wangAnonymousEfficientCertificateless2024}.

\subsection{Access Control and Privacy}
Access control and privacy preservation are among the most critical concerns when deploying CCN for 5G-based IoT systems. The inherent CCN characteristics of in-network caching and content name visibility exacerbate privacy risks and complicate traditional access control mechanisms.

Several lightweight access control mechanisms have been proposed to address these challenges. One example is Attribute-Based Encryption (ABE), which allows content producers to encrypt data based on an access policy defined by user attributes. Only consumers whose attributes satisfy the policy can decrypt the content. The work presented in \cite{mecerhedEfficientDecentralizedFinegrained2024}  introduces a lightweight variant of ABE tailored for CCN-based IoT systems. This variant features a hierarchical structure composed of multiple attribute authorities, each responsible for managing a subset of attributes and distributing the corresponding encryption keys. This decentralized design enhances scalability, allows for efficient attribute revocation, and enables the implementation of symmetric key exchange mechanisms, reducing computational overhead.

An alternative to traditional access control mechanisms is the use of decentralized approaches based on blockchain technology. In \cite{nourInformationCentricNetworkingWireless2021}, the authors introduce a decentralized authentication protocol for ICN-based 5G networks known as Blockchain-Based Decentralized Authentication Protocol (BDAP). It uses a distributed, immutable ledger to record authentication events, ensuring that only legitimate users can access protected content and network resources. Notably, the system is designed to operate without significant modifications to existing ICN-5G network infrastructure, thus promoting deployment and scalability in real-world environments.

Blockchain can be combined with Proxy-Based Re-Encryption (PBE) to enable fine-grained access control, as demonstrated in \cite{srikanthProxyBasedReEncryptionDesign2023}. PBE enables a semi-trusted proxy to transform ciphertexts from one user to another without accessing the underlying plaintext. Delegating computationally intensive cryptographic operations to a proxy is particularly advantageous in resource-constrained IoT environments, where end devices often lack the capability to perform complex encryption tasks. Furthermore, since the proxy does not need to be fully trusted, this approach simplifies trust management, enhances system scalability, and maintains data confidentiality.

In summary, securing CCN-based 5G IoT systems requires a multidimensional approach that is tailored to this architecture's unique characteristics. Cryptographic mechanisms must be lightweight yet effective, balancing performance with protection in environments with limited resources. Trust management enables secure, decentralized decision-making, allowing content and identities to be evaluated without relying on central authorities. Access control strategies, whether via attribute-based encryption or blockchain-enhanced delegation, must be scalable, granular, and privacy-preserving to match the dynamic, distributed nature of these systems. Table \ref{tab:overview_solutions} provides an overview of recent solutions that address these challenges.

Together, these security components form a cohesive framework that meets the evolving demands of the IoT, 5G, and CCN. This framework paves the way for secure and resilient next-generation communication systems.

\section{Emerging Trends and Future Research Directions}

As IoT, CCN, and 5G technologies converge, they are redefining the landscape of modern communication systems. New trends and research directions are emerging to address the growing need for security, scalability, and adaptability. Key emerging trends include:

\begin{itemize}
    \item \textbf{ML-driven security and network management:} 5G capabilities support ML-based security integration. As shown in \cite{patelAIBasedSecuritySystem2024}, ML enables functionalities like threat detection and automated remediation. In \cite{bukhowahDetectionDoSAttacks2024}, a DoS detection system using ML is proposed for IoT in CCN. ML facilitates rapid detection and response. Federated Learning (FL), in particular, is being explored to detect malware in IoT without centralizing data. The combination of FL with CCN-enabled IoT-5G supports decentralized intelligence while preserving privacy, aligning with CCN’s data-centric model \cite{reyFederatedLearningMalware2022}.
    \item \textbf{Zero Trust Architectures (ZTA) in distributed environments:} Traditional perimeter-based security is inadequate for dynamic, distributed systems. ZTA enforces “never trust, always verify” by embedding security into content rather than endpoints \cite{aleisa2025blockchain}. In CCN-5G settings, ZTA fits naturally. The model proposed in   \cite{ramezanpourIntelligentZeroTrust2022} combines ZTA with Artificial  (AI) for policy-based, context-aware access in 5G slices using Open RAN.
    \item \textbf{Blockchain and Distributed Ledger Technologies (DLTs) for trust management:} DLTs provide immutable, decentralized trust and complement CCN by securing content provenance, key management, and reputation systems without central authorities. Recent works integrate blockchain with IoT and edge computing for tamper-proof data sharing, smart contract-based access control, and trustless consensus \cite{usmanBlockchainBasedScalable2024}.
    \item \textbf{Post-Quantum Cryptography (PQC) for constrained devices:} As quantum computing advances, traditional cryptography (e.g., RSA, ECC) becomes vulnerable \cite{sharma2024qiotchain}. PQC provides quantum-resistant alternatives critical for CCN and IoT. In CCN, PQC is explored for content signing, secure naming, and key distribution \cite{sharmaPostQuantumCryptographyPQC2025}.
\item \textbf{Emerging technologies for 6G-IoT security and context awareness:} As research moves beyond 5G, 6G is expected to natively support intelligence, sensing, and global coverage, extending the capabilities of today’s IoT-5G systems. Several disruptive technologies will shape secure, intelligent 6G-IoT architectures. Non-Terrestrial Networks (NTNs) enhance coverage and resilience, especially in remote or infrastructure-less areas \cite{giordani2020non}. Virtual MIMO (V-MIMO) enables energy-efficient, high-capacity communications for large-scale IoT. THz sensing offers unprecedented resolution and bandwidth for context-aware applications, supporting adaptive security. Generative AI is emerging for proactive threat hunting, creating synthetic attack scenarios and improving predictive defense. Finally, explainable AI (XAI) \cite{arrieta2020explainable} enhances transparency and trust in ML-based security decisions, crucial for sensitive domains such as healthcare and autonomous systems.

\end{itemize}

The integration of AI/ML, Zero Trust Architectures, Blockchain/DLTs, and Post-Quantum Cryptography into the IoT-CCN-5G ecosystem marks a pivotal shift toward building resilient, intelligent, and secure communication infrastructures. These emerging trends tackle pressing issues such as distributed trust, data privacy, and post-quantum threats, while also enabling scalable, autonomous, and future-ready network architectures.

\section{Conclusion}
While the integration of the IoT, 5G, and CCN unlocks innovative applications, it also brings complex security challenges. In this paper, we examined the key security requirements for content-centric, 5G-based IoT systems and analyzed existing solutions.  We conclude that more research efforts should be made toward comprehensive solutions. Future work will focus on developing intelligent, context-aware security mechanisms for 5G-based, content-centric IoT systems.

\bibliographystyle{IEEEtran}
\bibliography{IEEEbib}

\end{document}